\documentstyle[12pt,epsfig]{article}
\input epsf
\begin{document}
\title{
   $J/\Psi $ suppression in heavy ion collisions
and the QCD phase transition
}
\author{ E.\ Shuryak, D. \ Teaney \\
   Department of Physics and Astronomy,\\
 State University of New York at Stony Brook,\\
 NY 11794-3800}
\date{\today}
\maketitle     

\begin{abstract}  
We suggest that the new regime  of $J/\psi $
suppression in Pb-Pb collisions found by  the NA50 experiment at CERN
is the result of non-trivial space-time
evolution due to specific behavior of the Equation of State (EOS)  
near the QCD phase transition.
We also study another suppression channel,
 the conversion of $J/\psi $ into $\eta_c$
during the late cool hadronic stage, and find it rather inefficient.
\end{abstract}   
\vspace{0.1in}
\centerline{\bf \large  SUNY--NTG 98-01}

\newpage  

1. Recent studies by the NA50 collaboration \cite{na50}
 in
Pb(158 GeV-A)-Pb collisions
 show   significant $J/\psi$ suppression compared
to an  extrapolation of known trends in lighter beams and p-A  
collisions. The data show a new suppression
mechanism for 
impact parameters  $b < 8 fm$. Surprisingly, this new suppression 
regime seems to set in very sharply.
Such abrupt behavior may be a  manifestation of  
dramatic  collective effects, related to qualitative
changes in  hot/dense hadronic matter due to the QCD phase transition.
\footnote{ We stress
however, that we discuss the new trend over the whole region 
and not just its onset.} 

Logically speaking, additional $J/\psi$ suppression can be 
explained either by  (i) increasing absorption rates  or 
by (ii) 
increasing the time that  $J/\psi$ spends in the hot/dense medium.
In contrast to the available literature devoted to the first scenario
we study the second one.
An increase in the Quark-Gluon Plasma(QGP) lifetime 
was found in the hydrodynamic framework in \cite{HS_sp}, as a result of a 
discontinuity in the Equation of State (EOS) at the QCD phase transition.

2. The oldest and the simplest argument for suppression
\cite{shu_78_photo} is 
that once QGP
  is formed, free gluons can rather easily
 ``ionize'' $J/\psi$ in an elementary process $gJ/\psi\rightarrow
 \bar c c$ similar to the photo-effect
 \cite{peskin,KhS_photo,KNS}. Later \cite{ms,KS}
 it was argued    that  successive charmonium levels
$\psi',\chi,J/\psi$ 
``melt''  at high temperatures and densities 
 due to Debye screening in the QGP,  
and ionizing gluons are superfluous.

%
%
%
%


There is a significant phenomenological difference between these two 
mechanisms. ``Melting'' implies the existence of certain thresholds as 
a function of matter energy density and can therefore generate a rapid
onset of suppression. 
Furthermore, it naturally explains why different charmonium states 
are suppressed at different densities. So, although there are
currently no
realistic models of this type, this scenario may be successful. 

Gluonic excitation on the other hand,  
 is simply a  probabilistic process,   
and like any other absorption processes (``co-movers'' for example)
cannot create discontinuous behavior by itself.
  Therefore  a recent paper \cite{KNS} proposed that 
the {\it formation} of QGP is discontinuous; QGP is produced
suddenly  after a certain critical energy
density $\epsilon_c$ is reached.
This work was motivated by the classical nucleation theory
of ``bubbles'' at a first-order phase transition and has a number of
theoretical and phenomenological problems. First, the thermodynamics
of the transition tells us that
plasma formation  is discontinuous in $temperature$ but not in
{\it energy density}. A mixed phase linearly interpolates between
the energy densities of the two phases,  
$\epsilon_{min}$ and  $\epsilon_{max}$. \footnote{The arguments of
\cite{KNS} rely on the non-equilibrium kinetics of bubble formation.
Note however, that a very small  
surface tension found by lattice studies, about $0.01 T^3_c$ \cite{tension},
 makes the formation of small bubbles quite probable 
\cite{Kapusta_etal}.}
Second, this scenario has a phenomenological problem:
 a  jump into the QGP phase at some $\epsilon_c$ changes  the
 equation of state discontinuously and implies a jump in 
entropy (or multiplicity)\cite{entropy}. Preliminary  data \cite{na50} 
on $<n_{ch}(E_T)>$ shows no discontinuity.

3. As deduced from lattice simulations at finite temperature, 
the QCD phase transition leads to a very non-trivial EOS. 
The ratio of pressure to energy density, $p/\epsilon$ 
has a deep minimum at the end of the  phase transition, $\epsilon_{max}
\sim 1-2 GeV/fm^3$. 
\footnote{
Roughly, since $p$ is the moving force and $\epsilon$ is the mass to 
be moved, acceleration is proportional to this ratio.} 
For the 
resonance gas and the high density QGP this ratio is relatively large (.2
and 1/3 respectively)  but is only about .05 at this
minimum. 
This minimum is referred to as the {\it softest point} of
the EOS and is associated with a slow hydrodynamic expansion. 
\cite{HS_sp,Rischke} 
When the initial energy density was close to the softest point, 
the QGP    
was found to live 2-3 times longer than the  more energetic and more central 
collisions at the SPS.\cite{HS_sp} 

In this letter we study the extent to which the non-trivial 
features of $J/\psi$ suppression can be explained by such an increase
in the QGP lifetime.
The hydro solution
in  \cite{HS_sp} was
made only for azimuthally symmetric central collisions.  
Assuming that about half of the collision energy is contained
in the fully stopped part of the matter, the above authors expect
$\epsilon_{max}$  to be reached at $P_{lab}\sim 40 GeV A$ for  
central AuAu collisions. (As this lies between the nominal AGS and SPS
energies, they proposed making additional low energy runs at the
SPS.) 
Of course, the initial energy density decreases with impact
parameter and since all values of $b$ are triggered on anyway,  
scanning $b$ is easier than scanning the  
collision energy in the experiment.
Examining the dependence of energy density on impact parameter, one
can see that indeed an impact parameter $b\approx 8 fm$ in Pb-Pb collision
roughly corresponds to the energy density of the observed change of
trend
in $J/\psi$ suppression.

Before turning to specifics, we point  out a qualitative difference
between our scenario and the one based on ``melting''.
In the latter case the various charmonium states are  
completely suppressed 
for $all$ $\epsilon>\epsilon_c$, while in our
case the suppression is centered  around $\epsilon\approx
\epsilon_{max}$ and the survival probability
increases for $\epsilon>\epsilon_{max}$. This difference is manifest in  
the specific 
predictions discussed below.

Hydro evolution for non-central collisions is much more complicated 
because of a directed flow (see
e.g.\cite{Olli,Sorge_elliptic}). Fortunately,   
we avoid these effects by restricting the discussion  
to $J/\psi$ suppression. For  
the first few fermi/c  
expansion is predominantly longitudinal, and thus
similar for central and non-central collisions.


4. It is clear that significant changes in the QGP lifetime should
affect $J/\psi$ suppression, even if absorption rates are unchanged.
However,
$J/\psi$ may escape from the system, which obviously limits the
sensitivity to QGP lifetime.

In order to reproduce many known observables, such as 
the distribution in
the transverse energy $E_T$ and its correlation with b,
we have constructed a small event generator to model
our scenario. For  Pb(158-GeVA)-Pb  
collisions at a
 given impact parameter, we generate charmonium events which
have survived nuclear absorption and are now in a hot de-confined
medium. The survival probability for a given event is  
$e^{-\Gamma\,t_p}$ where $t_p$ is the time spent transversing
this medium and $\Gamma$ is the ionization rate. $\Gamma$  
is chosen to match the observed 25\% suppression immediately
following the jump. Our goal is to provide a reasonable explanation
for the region near the discontinuity without too much regard for
the most central points.   

In our model $J/\psi$ interacts with the hot
medium only  
when the wounded nucleon density, $n_w$, is
greater than $3.1\,\mbox{fm}^{-2}$. This condition is similar to \cite{Blaizot}
and determines the transverse size of
 the region.
We estimate the longitudinal size from the Bjorken formula.   
The average energy density reached in a collision at impact parameter $b$
 is
\[
<\epsilon>= < \frac {3}{\Delta y\,c\,\tau} \frac{dE_T^o}{dA} >
= \frac {3\,q}{\Delta y\,c\,\tau}<n_w(b)>
\]
where $<n_w(b)>$ is the average wounded nucleon density, and
$q=.4\,GeV$ is a calorimeter dependent constant as defined in  
\cite{Kharzeev}. Since the jump is observed for
$b\simeq 8.5\,\mbox{fm}$ and  $<n_w(b= 8.5)>\simeq 2.6\,\mbox{fm}^{-2}$,
we choose $\tau= 1.9\,\mbox{fm}$ to   
 match the energy density of the jump to
 the softest point, $\epsilon_{max}= 1.6\,GeV/\mbox{fm}^3$. The thickness is  
then $2\,c\,\tau$, or the distance the colliding pancakes have
receded during the equilibration time.

In this model $J/\psi$ suppression begins only
for wounded nucleon densities that are
20\% larger than the softest point average, i.e, 
significant suppression begins
only above the mixed phase,
$\epsilon_{max}$ .
 For the lifetime of this hot region which destroys $J/\psi$ we take 
the forms shown in Fig.$\,$1a and calculate the corresponding suppression
patterns shown in Fig.$\,$1b .
To calculate the time inside plasma we need to distribute
these events in coordinate and momentum space.
The distribution in the transverse plane is given by Glauber theory,
like in
\cite{Kharzeev} and the distribution in the longitudinal direction is
presumed 
uniform.
We have calculated  transverse momentum distribution
of $J/\psi$  within Glauber theory (as in
\cite{Kharzeev2}). The longitudinal momentum or $x_f$ distribution is known
from p-A data and in the plasma frame follows the form   
$\sim(1-\left|x_f\right|)^{3.9}$. \cite{schuler,na38xf}

We  can then determine the time $t_p$ that $J/\psi$ remains in plasma and
the corresponding survival probability, $e^{-\Gamma\,t_p}$, and find
 the $J/\psi$ yield
\[
\frac{d^2\sigma^{J/\psi}}{db^2}
= 
\int_{\vec{s},z,\vec{p}_T,x_f}
\frac{d^2\sigma_{Glb}^{J/\psi}}  {db^2}
f_b(\vec{s}) f(z) f_{b,s}(\vec{p}_T) f(x_f) \, e^{-\Gamma\,t_p}   
\]
Here $\vec{s}$ is the coordinate in the transverse plane and each $f$ is  
a distribution functions described above. Such a $b$ dependent cross
section may be converted to an $E_T$ dependent cross section  
using the $E_T-b$ assignment of \cite{Kharzeev}.
Dividing this cross section by the Glauber value
we compare our survival probability   
with preliminary NA50 experimental results in Fig.$\,$1b .

The usual Glauber-type $J/\psi$ absorption on nucleons describes well  
p-A , lighter ion,  and peripheral Pb-Pb data and
and therefore we plot 
only 
$deviations$ from this trend in Fig.$\,$1b .  
The increased QGP lifetime
leads to additional suppression for intermediate $E_T$ 
between $50-100\,GeV$. 
Since $J/\psi$ may leak out of the plasma region the discontinuity
due to the lifetime is softened and we do not 
reproduce the very sharp jump seen in the data. However, 
agreement with most of the points
in the intermediate region is significantly improved. Furthermore,
our curves reproduce the vanishingly small
slope of the suppression in this region.
   
5. The $p_T$-dependence of the suppression may separate the different
suppression mechanisms discussed above. 
We have therefore calculated $<p_T^{2}>$ vs. $E_T$.  
As a function of $b$, $<p_T^{2}>$
is given by  
\[
<p_T^2>(\vec{b})= 
<p_T^{2}\,\frac {d^2\sigma^{J/\psi}} {d^2b}>   
\left/  
\frac {d^2\sigma^{J/\psi}} {d^2b} \right.  
\]
Converting the numerator and denominator to $E_T$ dependent cross
sections we plot $<p_T^{2}>$ vs. $E_T$ in Fig.$\,$1c .
As a result of the usual initial state parton re-scattering,
$<p_T^2>$ increases  with the centrality of the collision.
If  $J/\psi$ ``melt'' in the central region then
only peripheral $J/\psi$ survive, and  
$<p_T^2>$ decreases for the most central collisions, as suggested
by Kharzeev et al.     
In our scenario however, some $J/\psi$ may still ``leak'' even from the
central
regions, and this flattens out the $<p_T^{2}>\,vs.\,E_T$ dependence. 
This dependence
was indeed observed by the NA50 collaboration \cite{na50}.

\begin{figure}[ht]
\label{fig}
\begin{center}
\includegraphics[width=4.8cm]{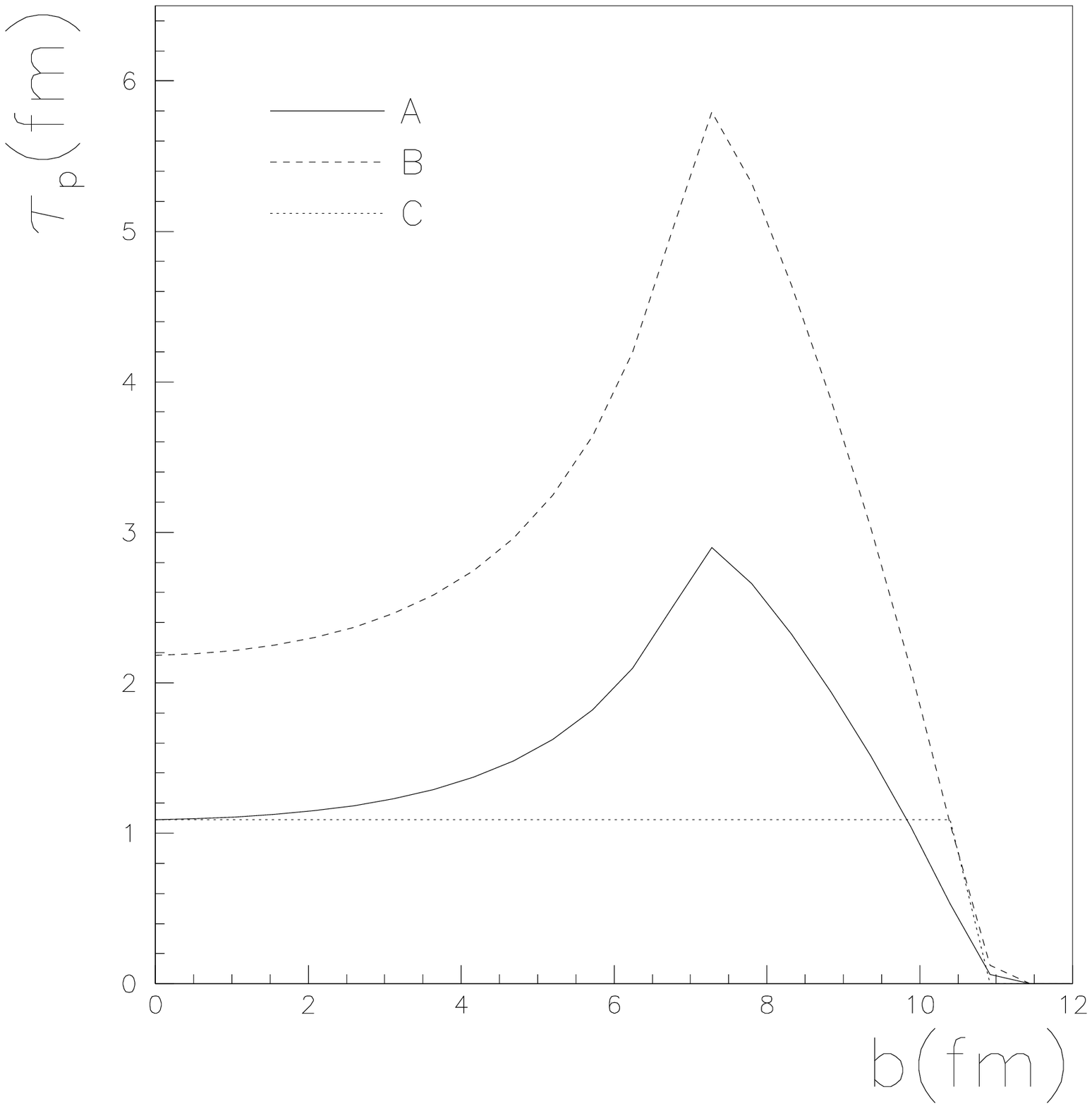}
\hspace{-.6cm}
\includegraphics[width=4.8cm]{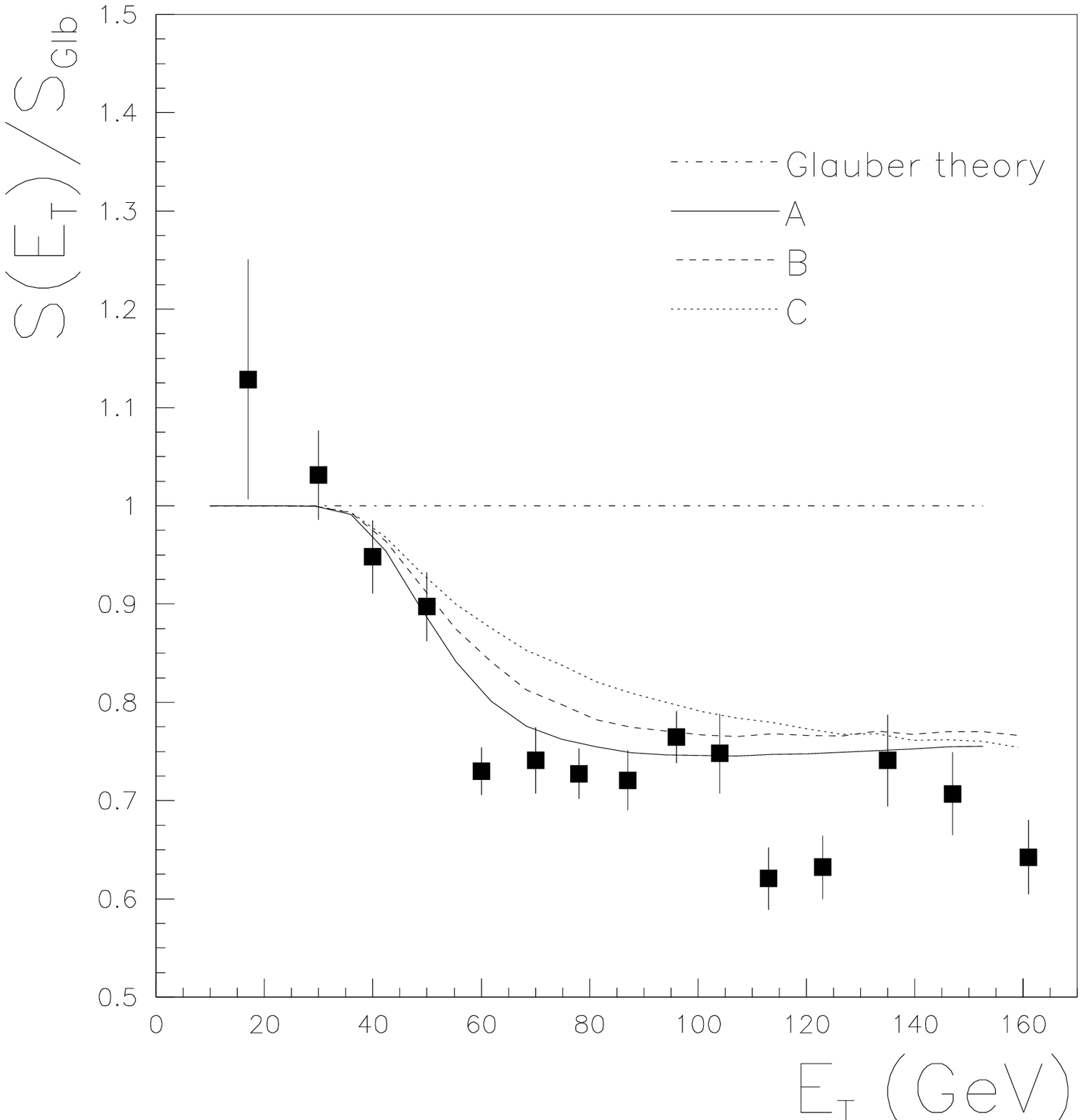}
\hspace{-.6cm}
\includegraphics[width=4.8cm]{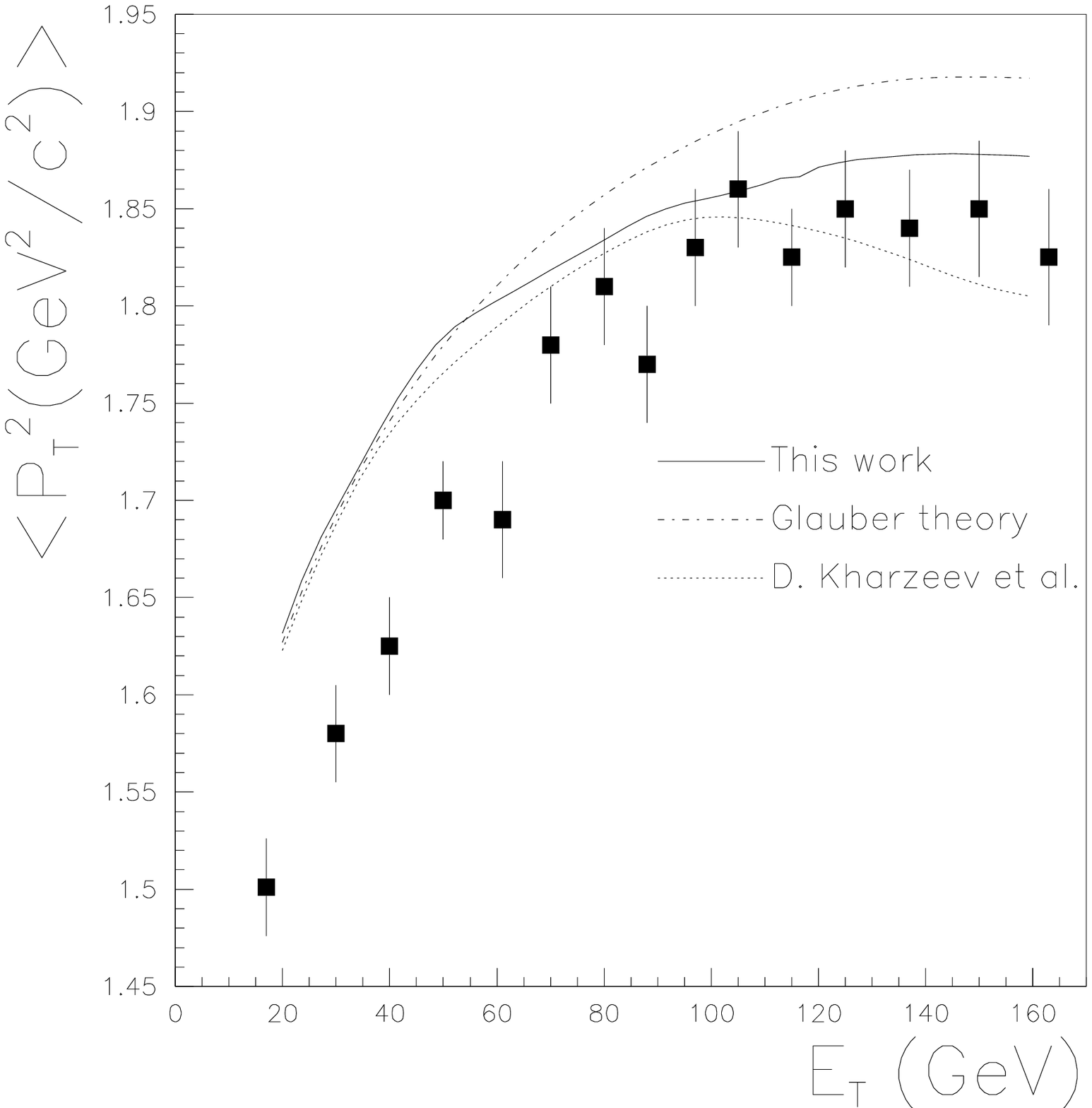}
\end{center}
\caption{
For the different lifetimes of the plasma region shown in (a) the
corresponding suppression pattern is shown in (b). (c)The $J/\psi$ $p_T$ distribution of this work 
(curve C in (a)) compared to $[16]$.}
\end{figure}

6.This part  is rather independent from the rest of the paper;  
Here we discuss  
the conversion of $J/\psi$ into its spin partner $\eta_c$.
This channel is one more potential ``sink'' for  
$J/\psi$, which (to our knowledge) 
has not been investigated previously.

These reactions may be important in the later stages of the 
collision. The  
``photo-effect'' above requires $\sim 1$ GeV gluons to ionize
$J/\psi$ and therefore can only proceed during the hot plasma stage  
of the collision. In contrast, since $J/\psi$ and $\eta_c$ are nearly
degenerate, spin exchange
reactions require little input energy
and may proceed during the cool hadronic stage, which lasts for
10-20 fm/c in Pb-Pb collisions. If these reactions have significant
rates, some fraction of $J/\psi$ can be converted into
$\eta_c$. The inverse reaction is of course also possible, but the  
rather short natural lifetime $\tau_{\eta_c}\simeq 15fm/c$ prevents  
significant feedback.   

The two
``spin exchange'' reactions we consider here are  $\psi +\pi\rightarrow  
\eta_c +\rho$ and $\psi +\rho\rightarrow \eta_c +\pi$. Below we 
calculate these reaction  rates in a hadronic gas at T=150$\,$MeV
 using a non-relativistic quark model.  

First we consider  the reaction $\psi +\pi\rightarrow\eta_c+\rho$. 
In a non  
relativistic quark model this reaction proceeds via a 
spin exchange collision   
between the c quark in  $\psi$ and the u quark in  $\pi$.  
We use a point like spin-spin interaction between the  
two particles  in the center of mass frame of the pion and $J/\psi$.
\begin{equation}
  H_{int}= C \vec{\sigma_c}\cdot\vec{\sigma_d}   
  \delta^3(\vec{r}_{\pi}-\vec{r}_{\psi})  
  \end{equation}
From   D-D$^*$ meson splitting 
\footnote{The exact physical nature of the spin-spin forces
between light and heavy quarks is not important for our results.  
Note however that it may be either due to the one-gluon
 exchange (as assumed in the works mentioned above)
or due to small size instantons, see S. Chernyshev, M.A. Nowak, I.
Zahed, Phys.\ Rev.\ D53,\ 5176\ (1996).}
we take the value  \cite{splittings}
 $C= 4.4\,(GeV)^{-2}$.
Using this interaction Hamiltonian
and assuming identical spatial wave functions for $\pi,\rho$ and $\psi,\eta_c$,
one obtains the following cross section  
%
%
%
\begin{equation}
	\sigma_{s\rightarrow s'}= \frac{C^2}{v_{\pi}+v_{\psi}}
	|\langle s'| \vec{\sigma}_c \cdot
		\vec{\sigma}_{\overline{q}} |s \rangle|^2
	\int \frac{d^3p_{\eta_c}}{(2\pi)^3} \frac{d^3p_{\rho}}{(2\pi)^3}
	(2\pi)^4\delta^4(p_{tot})
\end{equation}
The unpolarized cross section is found by summing the cross section over 
final spins and averaging over initial spins
\begin{equation}
	\sigma(E_{\pi})= 
	\frac{C^2}{\pi}
	\left[\frac{E_{\rho}p_{\rho}}{v_{\pi}+v_{\psi}}\right]_{cm}
	\langle\langle \sigma\cdot\sigma\ \rangle\rangle \\
\end{equation}
\[
	\mbox{  where  } \langle\langle \sigma\cdot\sigma \rangle\rangle
	=  \frac{1}{N_{s}}\sum_{ss'}
        |\langle s'| \vec{\sigma}_c \cdot
		\vec{\sigma}_{\overline{q}} |s \rangle|^2  \
\]
for $J/\psi + \pi\rightarrow\eta_c + \rho$ we have  
$\langle\langle\sigma\cdot\sigma\rangle\rangle= 1$ and for
$J/\psi+\rho\rightarrow \eta_c + \pi$ we have
$\langle\langle\sigma\cdot\sigma\rangle\rangle= 1/3$.  The value of the
 cross section(3) is  
 determined from the incident energy of the pion and the kinematics 
of the collision. For a stationary $J/\psi$ the  threshold for this 
process occurs when $E_{\pi}\simeq 710\,MeV$. The cross section is about 
1.2mb at $ 800\,MeV$.  

Now  we calculate the reaction rate in an ideal pion gas for  
the process
$J/\psi+\pi\rightarrow\eta_c+\rho$. The  
$J/\psi$ is at rest in the pion gas. In this energy range the pion mass  
is neglected and Boltzmann statistics are reasonable.
The reaction rate is given by (flux) x (cross section)
\begin{equation}
W_{\psi\rightarrow\eta_c}= 
	\frac{g_{\pi}}{2\pi^2}
	\int_{threshold}^{\infty}
	e^{\frac{-E_{\pi}}{T}}
	v_{\pi}E^2_{\pi}\sigma(E_{\pi})\,dE_{\pi}
\end{equation}
where $g_{\pi}= 3$ is the isospin of the pion. Using the cross section  
above (3) we have calculated the reaction rates numerically

\begin{equation}
W_{J/\psi + \pi\rightarrow\eta_c + \rho}= .29\,MeV  
\end{equation}
A very similar calculation gives the reaction rate for
$J/\psi+\rho\rightarrow\eta_c+\pi$. For example the unpolarized cross
 section
is given by $C^2/\pi^2\,[p_{\pi}E_{\pi}/(v_{\rho}+v_{J/\psi})]_{cm}
\langle\langle\sigma\cdot\sigma\rangle\rangle$. For this process we find
\begin{equation}
W_{J/\psi+\rho\rightarrow\eta_c+\pi}= .30\,Mev
\end{equation}

It has been suggested that the mass of the $\rho$ shifts downward 
in medium,
and the CERES dilepton data seem to support this claim.
 For reference we calculate the above rates for $m_{\rho}= 385\,MeV$
\begin{eqnarray}
W_{J/\psi + \pi\rightarrow\eta_c + \rho} \approx \\
W_{J/\psi + \rho\rightarrow\eta_c + \pi}  \approx 1.1\,MeV \nonumber
\end{eqnarray}
Even these rates are rather  small, and we conclude that it is
unlikely that this channel can kill more than few percent of $J/\psi$.

Finally, we consider experimental checks of $\eta_c$ production.
Its $\gamma\gamma$ branching is too
small to be useful. 
Among the known modes the most promising seems to be a $\phi\phi$
decay with branching ratio of .7\%.  
 As  the $\phi\phi$
background
is not large, $\phi$
identification/mass resolution in the KK channel
 is generally good, and the  $\eta_c$ peak   
is rather narrow,
$\eta_c$ identification may be feasible.

7.Discussion and Outlook. Recent findings by the NA50 collaboration 
indicate that an additional $J/\psi$ suppression
mechanism sets in for $b<8fm$. 
This suppression can be created by
a change in the absorption of $J/\psi$,  
or by a change in the space-time evolution, 
or both.
``Melting'' may  increase the absorption
of $J/\psi$ in plasma, while 
a change in the EOS may increase the lifetime of the plasma itself.


Two qualitative differences separate these scenarios experimentally.
First, ``melting'' is not directly related to the phase transition, 
but is a subtle phenomenon where basically $\chi$ states no longer
exist. The space-time evolution
on the other hand, {\it is} directly related to the  EOS and hence 
to the
transition point itself. Experimentally therefore, a change of 
 behavior at the corresponding impact parameter
\footnote{We mean discontinuous derivatives.
Above we criticized the scenario of sudden onset
of QGP because it produces discontinuous jumps in such global parameters
as entropy.}
in the hadronic observables sensitive to the transverse expansion  
, e.g. the $p_t$ slopes (especially of 
heavy secondaries N,d),  HBT transverse sizes etc., would signal 
a change in the space-time evolution of the system.

A second difference, 
is that while ``melting'' is presumed to destroy completely
the various charmonium states,  
the absorption in our scenario comes from the  usual 
probabilistic rates, and therefore $J/\psi$ may escape from the
central plasma region. This produces
a qualitative difference in the $mean$ transverse 
momentum   
vs. centrality, as discussed above. 
Much more detailed studies of  the $p_t$  dependence of
the suppression
are now underway and should clarify this issue.

  There remain some unresolved experimental issues with the
  preliminary data we use. 
    More  statistics and further analyses are needed to explain
some fluctuations in the data points and most importantly to
verify the shape of the transition to the new regime, 
which now appears as a discontinuous jump. Finally,
it is important to test experimentally an assumption, common to 
all explanations given above, i.e. that the phenomenon is  
induced at some value of {\it energy density}.   
This can be done by  lowering the collision energy of the Pb beam. 

8. Acknowledgments. We thank NA50 for allowing us to use their
preliminary
data, and 
D.Kharzeev and H.Sorge for helpful discussions.
This work is partially supported by US DOE grant DE-FG02-93ER40768.
\par

\end{document}